\documentclass[amsmath,amssymb,prb,letterpaper,
  twocolumn, floatfix, superscriptaddress,longbibliography]{revtex4-2}
\usepackage{bm,graphicx,hyperref}
\hypersetup{%
  breaklinks = {true},
  citecolor = {blue},
  colorlinks = {true},
  linkcolor = {red},
}

\begin{document}

\title{Layer-Resolved Quantum Transport in Twisted Bilayer Graphene: Counterflow and Machine Learning Predictions}

\author{Matheus H. Gobbo Kuhn}%
\affiliation {School of Engineering, Mackenzie Presbyterian University, S\~ao Paulo - 01302-907, Brazil}
\author{L. A. Silva}
\affiliation {School of Computing and Informatic, Mackenzie Presbyterian University, S\~ao Paulo - 01302-907, Brazil}
\author{D. A. Bahamon}
\email{dario.bahamon@mackenzie.br}
\affiliation {School of Engineering, Mackenzie Presbyterian University, S\~ao Paulo - 01302-907, Brazil}
\affiliation {MackGraphe - Graphene and Nanomaterials Research Institute, Mackenzie Presbyterian University, S\~ao
Paulo -01302-907, Brazil}

\date{\today}

\begin{abstract}
The layer-resolved quantum transport response of a twisted bilayer graphene device is investigated by driving a current through the bottom layer and measuring the induced voltage in the top layer. Devices with four- and eight-layer differentiated contacts were analyzed, revealing that in a nanoribbon geometry (four contacts), a longitudinal counterflow current emerges in the top layer, while in a square-junction configuration (eight contacts), this counterflow is accompanied by a transverse, or Hall, component. These effects persist despite weak coupling to contacts, onsite disorder, lattice relaxation and variations in device size. The observed counterflow response indicates a circulating interlayer current, which generates an in-plane magnetic moment excited by the injected current. Finally, due to the intricate relationship between the electrical layer response, energy, and twist angle, a clusterized machine learning model was trained, validated, and tested to predict various conductances.
\end{abstract}


\maketitle

The control of the relative angle between monolayers of two-dimensional materials is generating a new paradigm in the study of quantum materials \cite{Andrei:2021aa,doi:10.1021/acsnano.0c10435,Bernevig_Efetov_PT}. On one hand, different moir\'e materials have been investigated, such as twisted bilayer graphene (TBG) \cite{Cao18a,Cao18b,Cao19,Zhang18}, twisted multilayer graphene \cite{Cao19b,sctri_park,Lin22,Scammell_2022,Liu19}, and twisted transition metal dichalcogenides \cite{Wang_twTMC,Ghiotto_twTMC,Yang_twTMC,Hongyuan_twTMC,WangP_twTMC,Anderson_twTMC}, where topology and correlations have produced new electronic phases. On the other hand, the intrinsically chiral nature of twisted structures, that is, the lack of mirror symmetry, has led to the observation of large values of intrinsic ellipticity and circular dichroism (CD) controlled by the twist angle and not limited to the appearance of a magic angle \cite{Kim16}. 

The CD observed in TBG is associated with the emergence of an in-plane magnetic moment induced by the electric field of incident light. To understand the origin of this magnetic moment, it is crucial to consider the finite interlayer separation between the two graphene sheets \cite{Suarez_Morell_2017}, as well as the fact that currents in each layer can flow in opposite directions. This behavior is captured by the off-diagonal elements of the layer-resolved conductivity tensor, where $\sigma_{xy} = -\sigma_{yx}$ describes a Hall drag mechanism \cite{Suarez_Morell_2017}. This unconventional current distribution is not restricted to the transverse direction; Bistritzer and MacDonald \cite{Bistritzer11} have shown that at small twist angles, a longitudinal counterflow can also emerge. Investigations into the Drude weight in TBG have further demonstrated the existence of correlated currents between the layers, especially in the zero-frequency limit and under low electronic doping conditions \cite{Stauber18,Zhai23,PhysRevB.103.224436,PhysRevB.108.125415}. While these studies underscore the strong interlayer current correlations in TBG, they primarily focus on optical responses. Consequently, a significant open question remains: how does this interlayer current interdependence manifest in direct electrical transport measurements?

In this work, we address this question by investigating the electronic transport properties of nanoscale TBG devices featuring four- and eight-terminal configurations, corresponding to nanoribbon and square-junction geometries, respectively, with contacts selectively connected to individual layers. To characterize the layer-specific response, a current \( I \) is injected into the bottom layer (designated as the drive layer), and the resulting open-circuit voltage in the top layer (identified as the drag layer) is calculated. The interlayer current coupling is then quantified using the four-probe resistance \cite{Datta,DKFerry,StrongCoulombdrag,QuantumHalldrag,Helicaledgestates,InterlayerFQHE,StronglyCoupledExciton,FrictionalMagneto-CoulombDrag,PhysRevLett.60.2081,Buttiker4c,BEENAKKER19911,StoneResistance}. Our results reveal the emergence of a longitudinal counterflow current  in the drag layer when current is injected into the drive layer. In the square-junction setup, this  longitudinal counterflow is accompanied by a transverse, or Hall, counterflow. These phenomena remain robust in the presence of weak contact coupling, onsite disorder, lattice relaxation and variations in device size. Notably, the counterflow response reveals the presence of a circulating interlayer current, leading to the formation of an in-plane magnetic moment. This is consistent with observations from CD experiments in TBG \cite{Kim16}, but in our case, the effect is driven by the injected current rather than the electric field of light. Our findings demonstrate that exotic transport phenomena can be electrically probed in independently contacted layers, highlighting the crucial role of device–contact coupling in shaping the observed behavior.

The calculation of resistance as a function of Fermi energy and twist angle involves first determining the conductance matrix, followed by its inversion to solve for the voltages using a linear system \cite{Datta,DKFerry}. The computational cost is primarily driven by the calculation of the conductance terms. Although the number of terms can be reduced through symmetry, their values are strongly influenced by the interlayer coupling, which is mediated by the twist angle. Thus, obtaining the conductance matrix can be framed as a forward problem that can be tackled using machine learning (ML) techniques. These methods, particularly artificial neural networks, have shown great success in solving forward and inverse problems in photonics \cite{Piccinotti_2021,Ma2021,Jiang2021}, yet remain underutilized in quantum transport, where they could prove highly beneficial for complex systems like TBG. With this in mind, we use a Gradient Boosting Regressor (GBR) \cite{gbr1, gbr2, gbr3} to retrieve the conductance  $G_{p,q}$ as a function of the contacts  $p$, $q$, the Fermi Energy ($E_F$) and twist angle ($\theta$) ($G_{p,q} = f(p,q,E_F,\theta)$). Due to the complex relationship between the inputs and the conductance output, a single GBR  model does not  performed adequately. To address this, a divide-and-conquer strategy was implemented. In this approach, the data was first clustered, and then  GBR submodels were trained and validated within each cluster. The solutions from each cluster were finally combined to obtain \( G_{p,q} = f(p,q,E_F,\theta) \).


The paper is organized into two main parts. The first focuses on quantum transport. Section~\ref{sec:2C} examines a source-drain configuration with four terminals, highlighting the emergence of longitudinal counterflow. Section~\ref{sec:4C} analyzes the Hall-bar setup, emphasizing the transverse (Hall) response, while current distribution maps are presented in Section~\ref{sec:Imap}. The influence of disorder and other perturbations is discussed in Section~\ref{sec:disorder}. The second part introduces the ML approach in Section~\ref{sec:ML}, with the divide-and-conquer strategy detailed in Section~\ref{sec:MLDAC}.

\begin{center}
\begin{figure}[t]
\scalebox{0.85}{\includegraphics{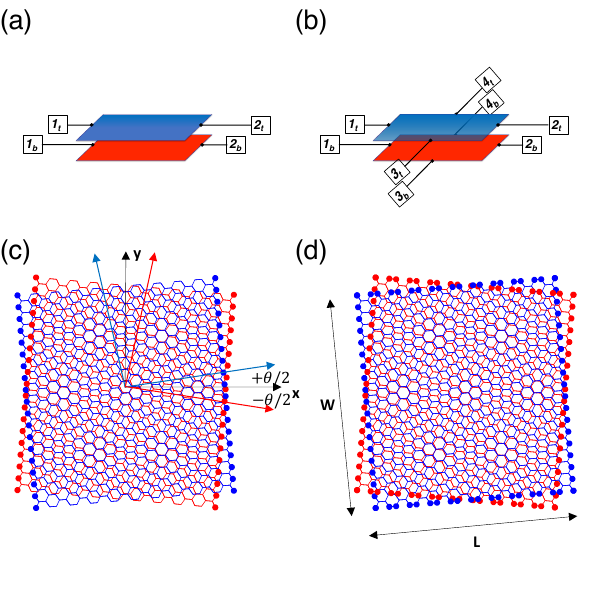}}
\caption{ (Color online) Scheme of the twisted bilayer graphene devices with layer-differentiated contacts for the four-contact device (a) and the eight-contact device (b). Atomic representation of the devices: (c) four-contact and (d) eight-contact. The filled dots correspond to the atomic sites in contact with the terminals.}
\label{fig:fig1} 
\end{figure}
\end{center}

\section{Model}
\label{sec:model}

A representation of the studied devices is shown in Fig. \ref{fig:fig1}(a)-(b), it consist of  one layer of graphene stacked atop other graphene layer. For positive twist angles, as shown in the figure, the top layer is rotated by $+\theta/2$ and the bottom layer by $-\theta/2$. For simplicity, in the device presentation, the contact appears at a single site, but in the atomic representation, it is connected to all the outermost atoms at the edges, as shown in the Fig. \ref{fig:fig1}(c)-(d). Unless otherwise specified, we will use a TBG region of dimensions \( W \times L \) with \( W = L = 50 \) nm. Initially, we attach the contacts to the zigzag edges of each layer of the central region, creating an armchair TBG nanoribbon. For the square-junction configuration, additional contacts are placed on the armchair edges of each layer. 

The Hamiltonian of the central region ($H_{TBG}$) is described by a tight-binding model where the hopping amplitudes between sites $i$ and $j$ are given by $ t_{ij}(d_{ij}) = V_{pp\sigma}(d_{ij})\cos^2(\phi) + V_{pp\pi}(d_{ij})\sin^2(\phi) $. Here, the bond length $ d_{ij} = |{\bm d}_{ij}| = |{\bm R}_j - {\bm R}_i| $, and $\phi$ denotes the angle formed by ${\bm d}_{ij}$ and the $z$-axis. It is important to note that the hoppings  depend on the bond length \cite{Brihuega12, Moon12}: $ V_{pp\sigma} = V_{pp\sigma}^0 e^{-\frac{d_{ij}-d_0}{\delta}} $ and $ V_{pp\pi} = V_{pp\pi}^0 e^{-\frac{d_{ij}-a}{\delta}} $, where $ V_{pp\sigma}^0 = t_{\perp}^0 = 0.48~\text{eV} $, $ V_{pp\pi}^0 = t_0 = -2.7~\text{eV} $, $ a = 0.142~\text{nm} $, $ d_0 = 0.335~\text{nm} $, and $ \delta = 0.184\sqrt{3}a $. Additionally, to accurately describe the electronic properties of TBG, for each site $i$, the neighbors $j$ are chosen within a disc of radius $ d_{ij} \leq 4a $.

To calculate the electrical response of the TBG devices, we start with the conductance between contacts $p$ and $q$, defined as $ G_{p,q} = \frac{2e^2}{h} \text{Tr}[\Gamma_p \mathcal{G} \Gamma_q \mathcal{G}^{\dagger}] $. In this expression, $\mathcal{G} = [E - H_{TBG} - \Sigma]^{-1} $ is the Green's function of the central region, and $\Sigma$ represents the self-energy terms, which are four for the TBG nanoribbon and eight for the TBG junction. Regardless of the number of terminals in the device, a wide band model is assumed for all of them. This means that the on-site term $\Sigma_c = -i|t|$, where $t$ is the nearest neighbor hopping parameter of graphene \cite{bahamon2020emergent,BahamonChV3}, is added to all atomic sites where the contacts are attached. These sites are represented in Fig. \ref{fig:fig1}(c)-(d) as filled dots. Equally, for each contact there is broadening function $\Gamma_{c} = i[\Sigma_{c} - \Sigma_{c}^\dagger]$.

\section{Two-terminal configuration with four contacts}  
\label{sec:2C}

\begin{center}
\begin{figure}[t]
\scalebox{0.85}{\includegraphics{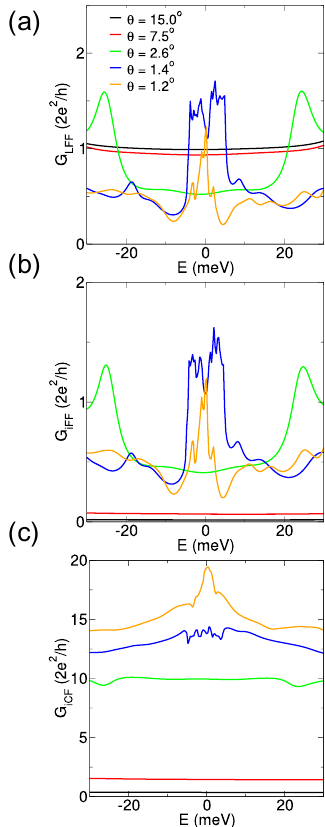}}
\caption{ (Color online) Independent terms of the conductance matrix of the TBG four-contact device: (a) Layer Forward Flow Conductance $G_{LFF} = G_{2t,1t} = G_{2b,1b}$, (b) interlayer Forward Flow Conductance $G_{iFF} = G_{2b,1t} = G_{2t,1b}$ and (c) interlayer Counter Flow Conductance $G_{iCF} = G_{1b,1t} = G_{2b,2t}$}
\label{fig:G4t} 
\end{figure}
\end{center}

We start with the central square TBG region  with  layer differentiated  leads attached to the source (left) and to the drain (right). The contacts attached to the left top ($t$) and bottom ($b$) layers are labeled as $1t$ and $1b$, while the contacts attached to the right are designated as $2t$ and $2b$. The layer differentiation allows us to treat the system as four-terminal device with  a relation between  current  and voltages at each terminal as  

\begin{equation}
\bf{I}
 = 
\left( 
\begin{array}{cccc}
G_{1t,1t}& - G_{1b,1t} & -G_{2t,1t} & -G_{2b,1t} \\
- G_{1b,1t} & G_{1b,1b} & -G_{2t,1b} & -G_{2b,1b}   \\
-G_{2t,1t}& -G_{2t,1b} & G_{2t,2t} & -G_{2b,2t}  \\
-G_{2b,1t} & -G_{2b,1b} & -G_{2b,2t} & G_{2b,2b}  \\
\end{array}
 \right)
\bf{V}.
 \label{eq:G4}
\end{equation}

\noindent In this expression the current vector is defined as ${\bf{I}} =\left( I_{1t}~I_{1b}~I_{2t}~I_{2b}\right)^T$ and the voltage vector ${\bf{V}} =\left( V_{1t}~V_{1b}~V_{2t}~V_{2b}\right)^T$. The diagonal terms of the conductance  matrix are $G_{1t,1t} = (G_{1b,1t} + G_{2t,1t} + G_{2b,1t} )$, $G_{1b,1b} =( G_{1b,1t} + G_{2t,1b} + G_{2b,1b})$, $G_{2t,2t} = (G_{2t,1t} + G_{2t,1b}  + G_{2b,2t})$; additionally, it is assumed that $G_{i,j} = G_{j,i}$. Based on the symmetry of the device, i.e., the same coupling to the leads of the top and bottom layer, only three independent conductance terms define the response matrix: (i) the Layer Forward Flow Conductance $G_{LFF} = G_{2t,1t} = G_{2b,1b}$, that is related to the probability that one electron injected into one layer exits at the opposite contact of the same layer; (ii) the interlayer Forward Flow Conductance $G_{iFF} = G_{2b,1t} = G_{2t,1b}$, that is associated with the probability of an electron injected into one layer forward flowing in the opposite layer; (iii) the  interlayer Counter Flow Conductance $G_{iCF} = G_{1b,1t} = G_{2b,2t}$ which pertains to the probability  of one electron injected into one layer counterflowing in the opposite layer.

In Figs. \ref{fig:G4t}(a)-(c), we can observe the three conductances for different twist angles.  For TBG, three coupling regimes between layers can be defined, dictating the lineshape of the conductance \cite{bahamon2020emergent,PhysRevLett.132.076302,PhysRevB.107.045418}. For large angles ($\theta \ge 10^{\circ}$), the layers are decoupled, resulting in a plateau at low energies with $G_{LFF}\sim (2e^2/h)$, while the values of $G_{iFF}$ and $G_{iCF}$ are low. As the twist angle is reduced ($2^{\circ} \le \theta < 10^{\circ}$), the coupling intensity increases, reducing $G_{LFF}$ and increasing the interlayer conductances such as $G_{iFF}$ and $G_{iCF}$. Finally, for small angles, the layers are strongly coupled, with $G_{LFF}\approx G_{iFF}$, exhibiting peaks \cite{PhysRevB.109.085412,PhysRevResearch.4.043145,ciepielewski2024transport} related to the high Density of States (DOS), entering the regime of magic angle \cite{BahamonChV3} when only one peak is observed. Remarkably, $G_{iCF}$ display the largest values of around $20\,(2e^2/h)$.

\subsection{Layer resistance}
\label{sec:2cLR}

To examine the layer-resolved response of our TBG device when a current \( I \) is injected into the  drive (bottom) layer, we use the four-probe resistance \( R_{mk,jn} = (V_j - V_n)/I \), which measures the voltage between terminals \( j \) and \( n \) when a fixed current is applied between terminals \( m \) and \( k \) \cite{PhysRevLett.60.2081,Buttiker4c,BEENAKKER19911,StoneResistance}. Operationally, we set $I = I_{1b} = -I_{2b}$, $I_{1t} = I_{2t} = 0$, and $V_{2b} = 0$, which enables us to truncate the fourth row and fourth column of the conductance matrix in Eq. \ref{eq:G4} \cite{Datta}. Subsequently, the $3\times3$ conductance matrix ($\tilde{G}$) is inverted to derive the resistance matrix, facilitating the determination of the voltages via

 \begin{equation}
\left( 
\begin{array}{c}
       V_{1t}\\
       V_{1b}\\
       V_{2t}
\end{array}
 \right) = \left( 
\begin{array}{ccc}
R_{1t,1t} & R_{1t,1b} &  R_{1t,2t} \\
R_{1b,1t} & R_{1b,1b} &  R_{1b,2t}\\
R_{2t,1t} & R_{2t,1b} & R_{2t,2t}
\end{array}
 \right)
 \left( 
\begin{array}{c}
       0\\
       I\\
       0
\end{array}
 \right).
 \label{eq:R3}
\end{equation}

\begin{center}
\begin{figure}[t]
\scalebox{0.85}{\includegraphics{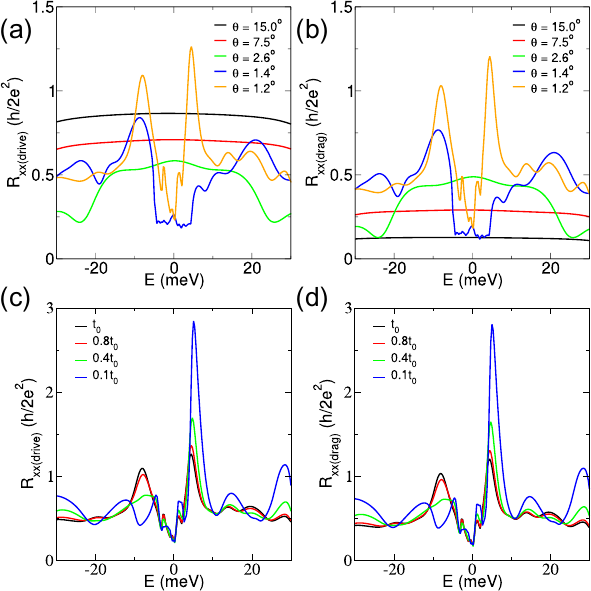}}
\caption{ (Color online) Four-contact TBG device: (a) Longitudinal resistance of the drive layer, $R_{\text{xx(drive)}}$, and (b) longitudinal resistance of the drag layer, $R_{\text{xx(drag)}}$.  
(c) $R_{\text{xx(drive)}}$ and (d) $R_{\text{xx(drag)}}$ for $\theta = 1.2^{\circ}$, illustrating the effect of varying the coupling between the contacts and the device using $t_0$, $0.8t_0$, $0.4t_0$, and $0.1t_0$, where $t_0$ represents the hopping energy between nearest neighbors.}
\label{fig:R4t} 
\end{figure}
\end{center}

The response of the drive  layer is accessed by measuring the voltage between the same terminals used to inject the current. From Eqs. \ref{eq:G4} and \ref{eq:R3}, we can write $ R_{1b2b,1b2b}= R_{1b,1b} = (G_{1t,1t}G_{2t,2t} - G_{2t,1t}^2)/|\tilde{G}|$, where $|\tilde{G}|$ is the determinant of $\tilde{G}$ \cite{Baranger4c,PhysRevB.46.9648}. To simplify the notation, we denote \( R_{1b2b,1b2b} \) as \( R_{\text{xx(drive)}} \). Fig.~\ref{fig:R4t}(a) displays the calculated  \( R_{\text{xx(drive)}} \) for various twist angles of the TBG. For large twist angles ($\theta \ge 10^{\circ}$), a constant resistance is observed, which decreases as the twist angle increases. For intermediate angles ($2^{\circ} \le \theta < 10^{\circ}$), the resistance continues to shrink, but some oscillations begin to appear. Finally, for low twist angles ($\theta < 2^{\circ}$), two broad peaks are observed around the charge neutrality point (CNP). It is worth noting that the layer resistance 

\begin{equation}
R_{\text{xx(drive)}} = \frac{(G_{\text{LFF}} + G_{\text{iFF}} + G_{\text{iCF}})^2 - G_{\text{LFF}}^2}{|\tilde{G}|},
\label{Eq:genRl}
\end{equation}

\noindent depends on various conductance terms, all of which contribute to the effective resistance of the layer.

\subsection{Counterflow resistance}
\label{sec:2cCF}

The response of the drag (top) layer is assessed through the resistance \( R_{1b2b,1t2t} = (V_{1t} - V_{2t})/I \), which relates the voltage drop across the top layer to the current flowing in the bottom layer. As with the drive layer, this quantity is denoted for simplicity as \( R_{\text{xx(drag)}} \). In Fig. \ref{fig:R4t}(b), the drag layer resistance is plotted for different twist angles of the TBG.  Firstly, it is important to mention that the generalized resistance clearly demonstrates a response from the drag layer to the current flow in the drive layer, regardless of the twist angle. Secondly, the resistance calculated for the different angles is always positive. In order to understand this result we have to note that $R_{\text{xx(drag)}} = \frac{G_{1b,1t}G_{2b,2t} - G_{2b,1t}G_{2t,1b}}{|\tilde{G}|}$. Or,  in other words, 

\begin{equation}
R_{\text{xx(drag)}} = \frac{G_{\text{iCF}}^2 - G_{\text{iFF}}^2}{|\tilde{G}|}.
\label{Eq:genRnl}
\end{equation}

\noindent This indicates that the generalized resistance is positive because the counterflow component exceeds the interlayer forward flow. And, it can be concluded that the current in the drag layer flows in the opposite direction to the drive layer. However, the behavior of $R_{\text{xx(drag)}}$ requires closer examination to determine whether it represents a genuine counterflow throughout the entire top layer or arises from the influence of effective contacts, which may cause significant electron scattering from contact $1b$ to contact $1t$.

Keeping that in mind, the coupling between the contacts and the central TBG region is reduced from the strong coupling regime (calculations presented so far), where the hopping between the contacts and the central region is \( t_0 \), to the weak coupling regime with a hopping strength of \( 0.1t_0 \), where \( t_0 \) represents the nearest-neighbor hopping energy.  Without loss of generality, the effect of coupling reduction on $R_{\text{xx(drive)}}$ and $R_{\text{xx(drag)}}$ is shown for the system with \( \theta = 1.2^\circ \) in Fig. \ref{fig:R4t}(c) and (d), respectively. While the coupling  significantly alters the resistance line shapes, minimizing external influences from the contacts highlights the intrinsic properties of the TBG region, resulting in an increase in drag resistance. Additionally, it is observed that the drive and drag resistances exhibit nearly identical responses. These findings suggest that the counterflow is not caused by scattering at the contacts but rather represents the natural response of the top layer to the current flow in the bottom layer. The results for both strong and weak coupling further demonstrate that \( R_{\text{xx(drag)}} \) exhibits a distinct behavior from the counterflow conductivity described in Ref.~\cite{Bistritzer11}, where the enhancement is attributed to the high DOS. In contrast, our results show a reduced \( R_{\text{xx(drag)}} \) in the high DOS regions compared to other areas.

\section{Hall-bar configuration with eight contacts}
\label{sec:4C}

Having established that a current in the drive layer induces a longitudinal flow in the drag layer, our attention now shifts to exploring the potential for inducing transverse and longitudinal flows \cite{Stauber18,Bistritzer11,Zhai23,PhysRevB.103.224436,PhysRevB.108.125415}. To achieve this, we introduce layer-specific contacts to the  armchair edges of  TBG. Similar to what was done in the previous case, the relation between the current ${\bf{I}} =\left( I_{1t}~I_{1b}~I_{2t}~I_{2b}~I_{3t}~I_{3b}~ I_{4t}~I_{4b}\right)^T$ and voltages  ${\bf{V}} =\left( V_{1t}~V_{1b}~V_{2t}~V_{2b}~V_{3t}~V_{3b}~V_{4t}~V_{4b}\right)^T$ at each terminal is determined by the conductance matrix ($\bf{I} = G V$).

Given the symmetries of the eight-terminal TBG device, out of the $8 \times 8$ elements of the conductance matrix, we have 10 independent terms. Note that despite having a square junction, our system does not possess a fourfold rotational symmetry \cite{Datta,Baranger4c,PhysRevB.46.9648}, which is due to the distinction between the armchair and zigzag directions \cite{doi:10.1021/nl201941f}. For clarity and focus, we do not present the full conductance values here. Nonetheless, we observe that quantum interference effects are suppressed due to the system being fully open. The three transport regimes defined by the interlayer coupling strength remain evident: as the twist angle decreases, interlayer conductance increases, while intralayer conductance decreases. Detailed conductance values for various twist angles are provided in Ref.~\cite{linkGBR}.

\subsection{Layer longitudinal resistances}
\label{sec:4cLR}

Although the dimensions of the involved matrices increase in the system of eight contacts compared to the one with four, the procedure remains identical. The contact injecting current and the one extracting it are identified. The voltage is set equal to zero for the latter, allowing the corresponding row and column to be eliminated from the conductance matrix. The reduced matrix $\tilde{G}$, with dimensions $7\times7$, is inverted to find the voltages at the other contacts. 

Since our focus is on the response of the layers, the current continues to be injected and extracted at the zigzag edges of the bottom layer. That is, $I_{1b} = I = -I_{2b}$ with $V_{2b} = 0$, and setting the currents to zero in the remaining contacts. The resistances in this setup are defined as \( R_{1b2b,jn} = (V_j - V_n)/I \), where \( j \) and \( n \) label any two contacts of the device.

We begin by evaluating the longitudinal resistance of the drive layer, defined as \( R_{\text{xx(drive)}} = R_{1b2b, 1b2b} \). At large twist angles, this resistance remains nearly constant. As the angle decreases, \( R_{\text{xx(drive)}} \) also decreases, exhibiting two distinct dips around the CNP, which gradually merge into a single dip in the magic-angle regime, as shown in Fig.~\ref{fig:RxH8c}(a). In contrast, the longitudinal resistance of the drag layer, \( R_{\text{xx(drag)}} = R_{1b2b,1t2t} \) shown in Fig.~\ref{fig:RxH8c}(b), increases as the angle decreases. For small angles, it also displays two dips that merge into one in the magic-angle regime.

It is important to highlight three key observations. First, similar to the four-terminal configuration, the longitudinal resistances of the drive and drag layers exhibit comparable magnitudes. Second, in this case, the resistance shows dips in the high DOS regions. Third, and most importantly, for all twist angles, \( R_{\text{xx(drag)}} \) remains positive. This indicates that the current injected at contact \( 1b \) couples more strongly to contact \( 1t \) than to \( 2t \), implying the emergence of a counterflow current in the drag layer.

\subsection{Layer Hall resistances}
\label{sec:4cHR}

\begin{center}
\begin{figure}[t]
\scalebox{0.85}{\includegraphics{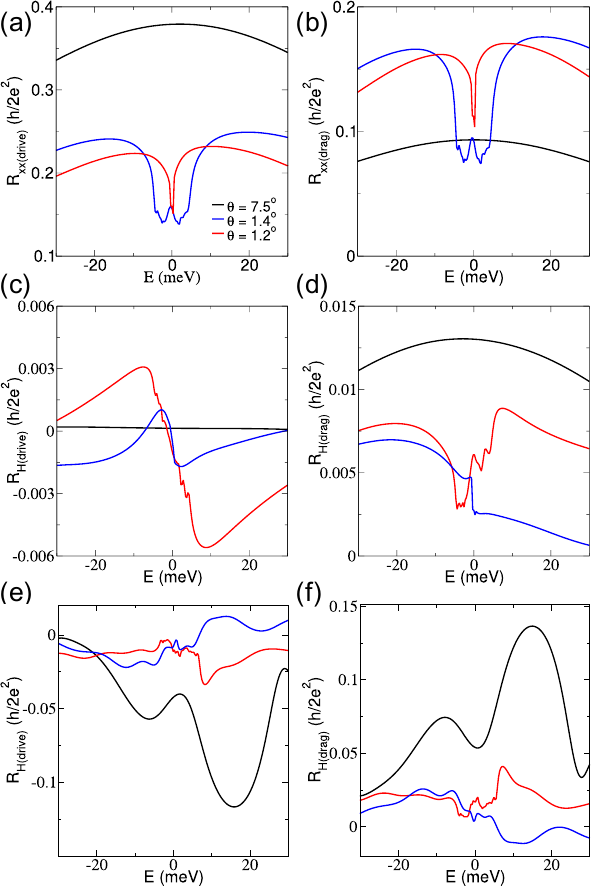}}
\caption{ (Color online) TBG device with eight-layer differentiated contacts: (a) \( R_{\text{xx(drive)}} \), (b) \( R_{\text{xx(drag)}} \), (c) \( R_{\text{H(drive)}} \), and (d) \( R_{\text{H(drag)}} \). Panels (e) and (f) show the same device under weak coupling (\( 0.1t_0 \)) between the contacts and the central TBG region, presenting \( R_{\text{H(drive)}} \) and \( R_{\text{H(drag)}} \), respectively.}
\label{fig:RxH8c} 
\end{figure}
\end{center}

We now focus on examining the accumulation of charge in the  transverse direction to the current flow \cite{Zhai23,PhysRevB.108.125415}, first in the  drive  layer ($ R_{\text{H(drive)}} = R_{1b2b, 4b3b} $) and then in the drag  layer ($R_{\text{H(drag)}}   = R_{1b2b, 4t3t} $). As shown in Fig. \ref{fig:RxH8c}(c) for the drive  layer, although the transverse charge accumulation is small, it increases when reducing the angle. A change in the resistance signal is also observed, indicating that the charge accumulation is reversed when passing through the CNP.

For the drag layer, the Hall resistance is 25 times greater than $ R_{\text{H(drive)}}$, which is noticeable even for large angles, as observed in Fig. \ref{fig:RxH8c}(d). The observation of this asymmetric charge accumulation is only possible by contacting each layer separately. In the absence of a magnetic field, a traditional setup would yield a Hall resistance of zero \cite{Datta}, in contrast to what is observed here. It is also important to  highlight that the direction of the layer Hall voltage can be reversed by changing the twist direction ($ R_{\text{H(drive)}}(\theta) = -R_{\text{H(drive)}}(-\theta)$)

To isolate contact effects, we  reduce the coupling to the central region to $0.1t_0$. Figs.~\ref{fig:RxH8c}(e) and (f) show the resulting Hall resistances $R_{\mathrm{H(drive)}}$ and $R_{\mathrm{H(drag)}}$ under these weak coupling conditions. Notably, both resistances increase across all twist angles, with the $7.5^\circ$ case (black curve) exhibiting particularly strong values despite the suppressed interlayer coupling. This behavior agrees with reported enhancements of chiral effects in TBG away from the magic angle \cite{Kim16}. The reduced device openness under weak coupling also reveals a clear dipole Hall signature \cite{Zhai23}, manifested as $R_{\mathrm{H(drive)}} \approx -R_{\mathrm{H(drag)}}$, regardless of the twist angle.

\subsection{Local current distribution}
\label{sec:Imap}

\begin{figure} [t]
\centering 
\includegraphics[width=1\linewidth]{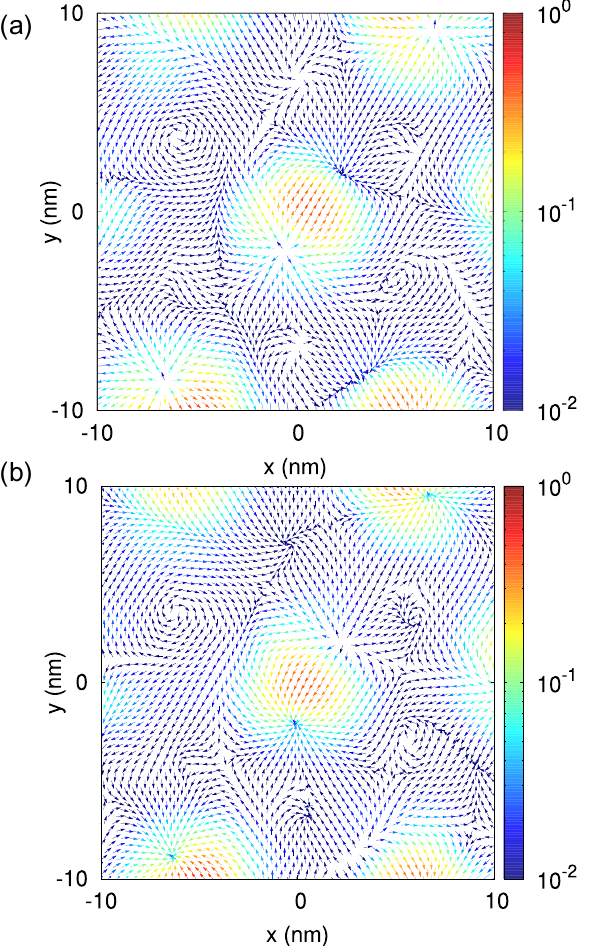} 
\caption{(Color online) Averaged current distribution in the central \(20 \times 20~\text{nm}^2\) region for (a) the drive layer and (b) the drag layer. Arrowhead colors indicate the magnitude of the current vectors, normalized to the maximum bond current, which occurs in AA-stacked regions.
}
\label{fig:Imap} 
\end{figure}

To gain deeper insight into the counterflow effect, we examine the bond current between sites \(i\) and \(j\), defined as $I_{ij} = \frac{2e}{h} \int \left[ t_{ji} G^<_{ij} - t_{ij} G^<_{ji} \right] dE$, where \(t_{ij}\) is the hopping energy between the sites and \(G^<_{ij}\) is the lesser Green's function~\cite{bahamon2020emergent,PhysRevB.80.165316}.
For ease of visualization of the current patterns—and without loss of generality—a bias of $10~\mu$eV was applied between the left ($1b$) and right contacts ($2b$) of the drive layer. The bond currents at each atomic site were then averaged over each hexagonal plaquette. To avoid double counting, the centers of the hexagons were chosen according to a triangular lattice with lattice constant \(3a\) \cite{bahamon2020emergent}.

The averaged current distribution is shown for the central \(20 \times 20~\text{nm}^2\) region in Fig.~\ref{fig:Imap}(a) (drive layer) and Fig.~\ref{fig:Imap}(b) (drag layer). The arrowhead colors represent the current vector magnitudes, normalized to the maximum bond current occurring in AA-stacked regions. Remarkably, these regions exhibit currents two orders of magnitude greater than the injected current at the \(1b\) contact.

The vector maps clearly demonstrate that despite complex current patterns, the two layers exhibit counterflow - with currents consistently flowing in opposite directions. While we focus on the central region, this behavior persists throughout the entire device, independent of local current magnitude or distribution complexity.

At the atomic scale, the current flow shows significant inhomogeneity even within the drive layer. The AA-stacked regions, for instance, display current vectors with varying orientations. We also identify distinct source and sink regions where current transfers between layers, with flow directions alternating between bottom-to-top and top-to-bottom.

\subsection{Disorder, relaxation and size effects} 
\label{sec:disorder}

Up to now, our results for TBG devices with four and eight contacts have demonstrated the induction of longitudinal and transverse open-circuit voltages in the drag layer as a result of an injected current in the drive layer. However, it is well established that mesoscopic current and voltage measurements are both sample-specific and non-local \cite{Datta,StoneResistance}. With this in mind, we will examine the influence of three types of perturbations. First, we will assess the effects of onsite disorder and lattice relaxation on a $50 \times 50$ nm$^2$ sample. Then, we will explore the impact of sample size by analyzing a pristine device with dimensions of $80 \times 80$ nm$^2$.

\begin{figure} [t]
\centering 
\includegraphics[width=1\linewidth]{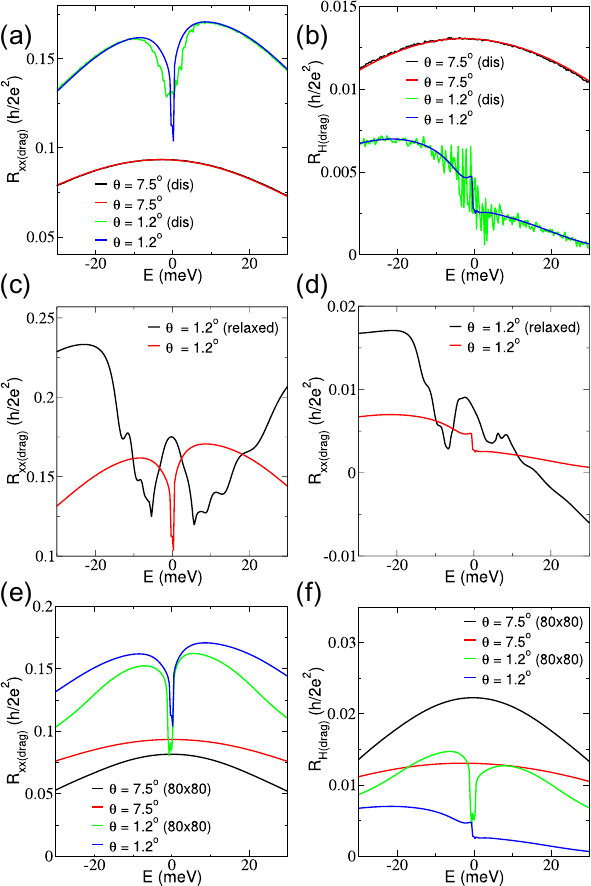} 
\caption{(Color online) Drag resistances under different perturbations: (a) \( R_{\text{xx(drag)}} \) and (b) \( R_{\text{H(drag)}} \) for the disordered TBG system (black and green curves), with the red and blue curves representing the pristine case. (c) \( R_{\text{xx(drag)}} \) and (d) \( R_{\text{H(drag)}} \) for the lattice-relaxed device, where the red curve corresponds to the unrelaxed case. For a pristine TBG device of size \(80 \times 80~\text{nm}^2\) (black and green curves), (e) \( R_{\text{xx(drag)}} \) and (f) \( R_{\text{H(drag)}} \) are shown, with the red and blue curves corresponding to a device of size \(50 \times 50~\text{nm}^2\).
}
\label{fig:Dis80} 
\end{figure}

For the disorder case, a low concentration of short-range impurities—occupying 10\% of the lattice sites—is randomly introduced by assigning onsite energies  selected from a uniform distribution within the range \([-0.1t_0, +0.1t_0]\). The conductance matrix elements are averaged over 10 different  configurations, and the linear system is then solved to obtain the  four-probe resistance. The results are presented in Fig. \ref{fig:Dis80}(a) for $R_{\text{xx(drag)}}$ and in Fig. \ref{fig:Dis80}(b) for $R_{\text{H(drag)}}$, indicating that disorder has a minimal impact on the strength of the induced longitudinal and Hall counterflows. For a large twist angle (\(\theta = 7.5^{\circ}\)), the disordered resistance closely resembles that of the pristine case, with both curves nearly overlapping. In contrast, for a low twist angle (\(\theta = 1.2^{\circ}\)), the high DOS near the CNP allows electrons to scatter into multiple states with the same energy, leading to a broadening of the dip in $R_{\text{xx(drag)}}$ or oscillations in \(R_{\text{H(drag)}}\). Despite these disorder-induced modifications, the fundamental counterflow behavior remains preserved, reinforcing the robustness of the effect across different disorder configurations and twist angles.

In the case of lattice relaxation, it is important to first recall that for twist angles larger than  $2^\circ$, the atomic structure remains nearly undeformed, and the impact on the electronic properties is minimal. In contrast, below $2^\circ$ significant structural reconstruction takes place. The AA-stacked regions shrink, while the AB/BA-stacked regions expand. This relaxation effect increases the energy gap between the low-energy minibands and enhances the Fermi velocity \cite{Nam17}.

With this in mind, we implemented an in-plane lattice relaxation model \cite{Nam17,Nam17Err} for $\theta = 1.2^\circ$. The resulting  $R_{\text{xx(drag)}}$ is shown in Fig.~\ref{fig:Dis80}(c), and the corresponding $R_{\text{H(drag)}}$ is presented in Fig.~\ref{fig:Dis80}(d). These results clearly show that lattice relaxation enhances the calculated resistance values. This can be understood from three perspectives. First, the counterflow current is not confined to the AA-stacked regions, as evidenced by the current distribution map in the previous section. Second, relaxation modifies the interlayer potential, which can promote increased interlayer current flow. Third, although relaxation may break certain lattice symmetries, the atomic displacements rotate around the centers of the AA regions~\cite{Nam17}. As a result, when the twist angle is reversed, the displacement vectors also reverse, and the overall structure retains its chirality.

To investigate the effect of the central region's size, the four-probe resistances were calculated for a TBG device with dimensions \(W \times L\), where \(W = L = 80\) nm. The longitudinal drag resistance \(R_{\text{xx(drag)}}\), shown in Fig. \ref{fig:Dis80}(e), exhibits a response similar to that of the smaller \(50 \times 50\) nm\(^2\) device. In contrast, \(R_{\text{H(drag)}}\),  presented in Fig. \ref{fig:Dis80}(f), maintains a comparable strength but displays a different line shape, which is particularly evident for low twist angles. Since our calculations consider fully coherent transport in an entirely open system, electrons in the TBG interact with the entire device, including the highly doped contacts, making it unreasonable to expect identical response shapes \cite{StoneResistance,Buttiker4c}. The key takeaway here is that, even in a larger system, the counterflow response remains robust.

\section{Machine Learning approach}
\label{sec:ML}

From the strictly scientific point of view, we have shown the appearance of longitudinal and Hall counterflow in the drag layer. Our analysis relies on the numerical calculation of the conductance matrix  using the Green's function, which is the most computationally intensive part, and its subsequent  inversion. With the idea of developing and evaluating new numerical tools to access the transport properties of devices, in this section we show our effort to calculate the four probe resistances of  TBG using ML techniques.

The key benefit of using ML is the substantial reduction in computational cost compared to traditional simulations \cite{cc, cc2, cc3}, without sacrificing accuracy. Once the ML model is trained, it can quickly compute the individual terms of the conductance matrix (\( G_{p,q} = f(p,q,E_F,\theta) \)), which are then incorporated into the conventional quantum transport framework to calculate the resistance \(R_{mk,jn}\). This hybrid approach significantly reduces the computation time by addressing the most resource-intensive part of the process. With this in mind the  Gradient Boosting Regressor (GBR) \cite{gbr1, gbr2, gbr3} was trained, validated, and tested.

To generate inputs for the ML models, over 289,000 samples were collected. Each sample contains the input electrical contact (injection lead), the output electrical contact (extraction lead), the normalized energy, the twist angle, and the normalized conductance. The samples span energy values ranging from approximately \(-40\) meV to 40 meV, with an energy step of 0.3 meV. A total of 18 different twist angles were considered, ranging from \(1.17^\circ\) to \(4.0^\circ\), with a higher density of samples between  \(1.20^\circ\) and  \(1.30^\circ\). This results in about 13,200 samples for each twist angle in the dataset. 

Normalization was applied to standardize their values around a reference point, ensuring a consistent numerical range across all angles and making the input data comparable for the ML models \cite{datamining}. Specifically, the first normalization was done using Scikit-learn's \texttt{StandardScaler} \cite{pedregosa2011scikit, normalização1} function, adjusting the data to have a mean of zero and a standard deviation of one. The second normalization was carried out with the \texttt{MinMaxScaler} \cite{pedregosa2011scikit, normalização1} function, also from Scikit-learn, scaling the data to a range between zero and one. These steps are essential to ensure that the input variables are comparable in scale and magnitude \cite{normalização1}, preventing the models from being unduly influenced by differences in scale \cite{datamining}.

\subsection{ Divide-and-conquer} 
\label{sec:MLDAC}

\begin{center}
\begin{figure} [t]
\centering 
\includegraphics[width=1\linewidth]{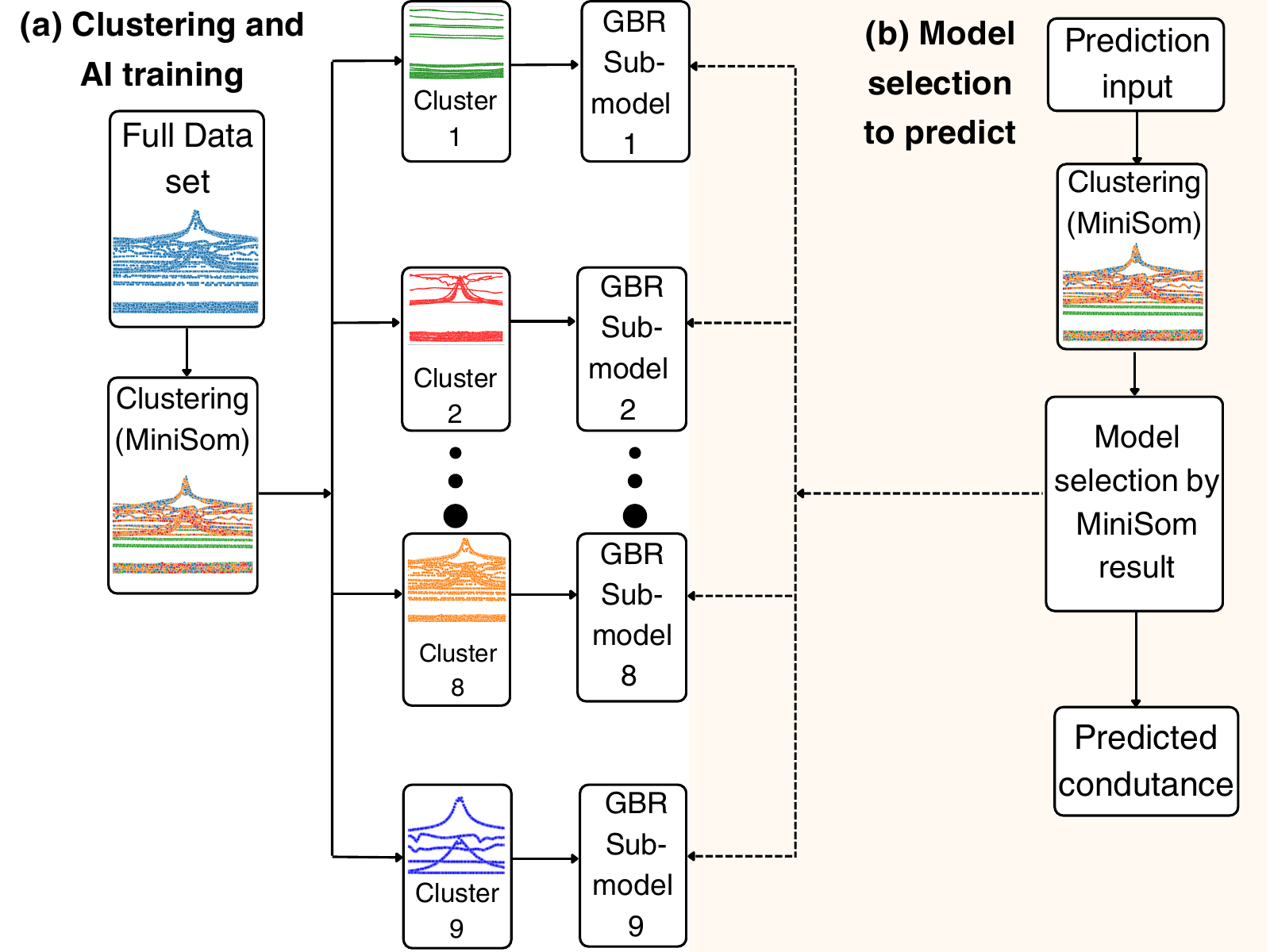} 
\caption{Data flowchart illustrating the clustering, training, and prediction procedures.} 
\label{fig:Diagrama} 
\end{figure}
\end{center}

To address the complexity of predicting the 56 off-diagonal elements of the conductance matrix, MiniSom \cite{mini1, som2} was employed to segment the data into clusters as the first step of a divide-and-conquer strategy. Fig. \ref{fig:Diagrama} illustrates the entire process. Initially, the data were segmented into clusters, and then independent GBR submodels  were trained for each cluster \cite{trivedi2015utility, cp1, cp2, cp3, cp4,cp5}. This is demonstrated in panel (a) ``Clustering and AI Training," where the segmentation of the data and subsequent training of AI submodels are shown. The second part, in panel (b) ``Model Selection to Predict", details the automated process of selecting the most appropriate submodel for conductance prediction based on the inputs. These inputs are processed by MiniSom, which identifies the corresponding cluster, after which the GBR sub-model trained for that specific cluster is selected to make the prediction.

To identify the optimal number of clusters, we applied two methods: the Akaike Information Criterion (AIC) and the Bayesian Information Criterion (BIC) \cite{Sowan2023, pedregosa2011scikit, Bulteel2013}. These criteria aim to determine the number of clusters that minimize their respective values, ensuring a balance between model simplicity and data fit. Both AIC and BIC were evaluated across a range of 1 to 12 clusters, and notably, both methods identified 9 clusters as the optimal choice.

\subsection{Hyperparameter Tuning, Model Training, and Validation}

For each GBR submodel, the tuned hyperparameters include the learning rate (\(\lambda\)), the number of estimators (\(N_{\text{estimators}}\)), and the maximum tree depth (\(max\_depth\)), as these parameters play a crucial role in shaping the model’s learning and generalization abilities. All other parameters are kept at their default values \cite{manualGBR}. \(\lambda\) determines the impact of each tree on the final prediction, while \(N_{\text{estimators}}\) specifies the number of boosting stages. Meanwhile, the maximum of \(max\_depth\) sets a limit on the number of nodes in each tree, controlling how deep the trees can grow before stopping \cite{manualGBR}.

\begin{figure}
    \centering
    \includegraphics[width=0.47\textwidth]{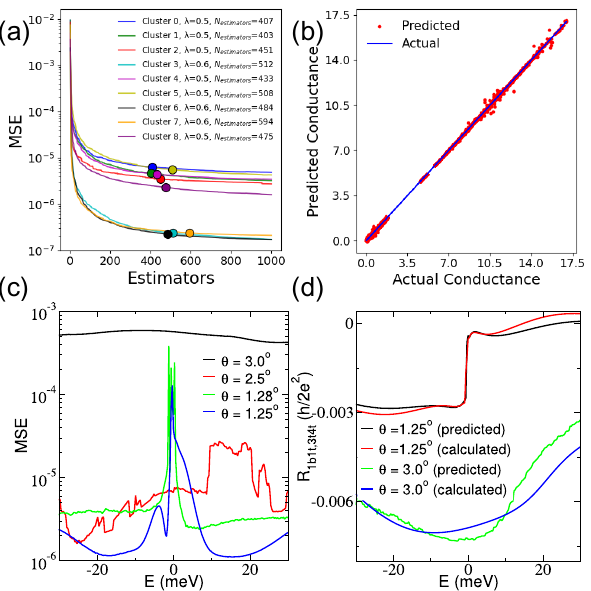}
    \caption{(a) Validation error plots for each cluster, showing the selected $N_{\text{estimators}}$ values for each GBR submodel. The validation error curve was used to determine the optimal number of estimators, with red circles marking the selected points. (b) Comparison of predicted and actual conductance values. (c) MSE of the conductance predictions. (d) Resistance predictions from the GBR model are compared to results obtained via Green's functions for twist angles not included in the training and validation datasets.}
    \label{MSEGBR}
\end{figure}

The tuning process is structured into three stages \cite{manualGBR}: data preparation, searching for the optimal combination of \(\lambda\) and \(N_{\text{estimators}}\), and finally, optimizing tree depth.  In the first stage, the dataset is randomly split, with 80\% allocated for training and 20\% for validation. The second stage involves fixing \(max\_depth\) at 5 while training individual submodels using various combinations of \(\lambda = \{1.0, 0.7, 0.6, 0.5, 0.4, 0.3\}\) and \(0 < N_{\text{estimators}} \leq 1000\). To guide hyperparameter selection, the mean squared error (MSE) is computed at each boosting iteration. Training is stopped when the MSE stabilizes or begins to increase, indicating that the model has reached its optimal generalization point before overfitting occurs \cite{pedregosa2011scikit, manualGBR}. It is important to note that, since multiple combinations of (\(\lambda\), \(N_{\text{estimators}}\)) can yield similar MSE values, configurations that achieve low error while minimizing the number of estimators are prioritized \cite{manualGBR}. Fig. \ref{MSEGBR}(a) illustrates this process for the nine GBR submodels, where the MSE curves are analyzed to determine $N_{\text{estimators}}$ for the $\lambda$ values shown in the legend. The chosen values for each cluster are represented by colored dots and indicated in the legend.  In the third stage, after determining the optimal \(\lambda\) and \(N_{\text{estimators}}\), the tree depth is refined by evaluating the MSE for different \(max\_depth\) values ranging from 5 to 20. The guiding principle in this step is to choose the smallest tree depth that minimizes the MSE \cite{manualGBR}.

The final set of selected hyperparameters is summarized in Table~\ref{tab:GBR_params}. These values define the architecture of the trained GBR submodels, which are subsequently used for conductance predictions. The performance of these predictions on the validation dataset is illustrated in Fig.~\ref{MSEGBR}(b), where the blue line represents the actual conductance values, and the red dots indicate the predictions made by the GBR model. The strong agreement between the predicted and actual values confirms that the model effectively captures the underlying physical relationships governing conductance estimation. 

As an additional test, we assess the model's predictive performance for twist angles not present in the training or validation datasets. Fig.~\ref{MSEGBR}(c) depicts the MSE, defined as \(\text{MSE} = \frac{1}{N} \sum_{i,j} (G^p_{i,j}(E,\theta) - G^c_{i,j}(E,\theta))^2\), as a function of energy for these previously excluded twist angles. Here, $N=56$ represents the number of off-diagonal terms, and $G^k_{i,j}$ denotes the conductance predicted ($k=p$) by the GBR model or computed ($k=c$) using Green's functions. Overall, the observed MSE values remain low, but certain features highlight the strong dependence of transport properties in TBG on the twist angle. For \(\theta = 3.0^\circ\), the black curve stays relatively flat across the energy range, maintaining a higher and nearly constant value. This indicates the absence of significant features or sharp variations in conductance, in contrast to the lower-angle cases, where a pronounced peak at \(E = 0\) meV emerges due to high DOS.

\begin{table}[b]
    \centering
    \caption{Selected GBR hyperparameters for each cluster.}
    \label{tab:GBR_params}
    \begin{tabular}{c c c c}
        \hline
        Cluster & $\lambda$ & $N_{\text{estimators}}$ & \(max\_depth\) \\
        \hline
        0 & 0.5 & 407 & 10 \\
        1 & 0.5 & 403 & 9 \\
        2 & 0.5 & 451 & 6 \\
        3 & 0.6 & 512 & 15 \\
        4 & 0.5 & 433 & 13 \\
        5 & 0.5 & 508 & 9 \\
        6 & 0.6 & 484 & 10 \\
        7 & 0.4 & 593 & 7 \\
        8 & 0.5 & 475 & 8 \\
        \hline
    \end{tabular}
\end{table}

These predictive limitations are reflected in the four-probe resistance, as shown in Fig.~\ref{MSEGBR}(d), where the predicted \( R_{1b1t,3t4t} \) is compared to the calculated values. For \(\theta = 3.0^\circ\), both curves follow a similar overall trend, featuring a wide dip. However, the predicted resistance appears noisier and alternates between underestimating and overestimating the calculated values across the entire energy range.  For \(\theta = 1.25^\circ\), both curves exhibit strong agreement, with the GBR model successfully capturing the step discontinuity around the CNP.  From a quantum transport perspective, it is important to emphasize that this resistance has not been previously presented. It was selected both to evaluate the predictive capability of the GBR model and to underscore the complexity of layer-resolved quantum transport phenomena in TBG. 

\section{Final remarks}


This work is structured into two main, well-defined yet interconnected sections. The first section presents and analyzes the layer-resolved quantum transport response of TBG using a traditional method. The second section, which relies heavily on data generated from Green’s functions, introduces a novel divide-and-conquer ML approach to compute all elements of the conductance matrix. Together, this work advances our understanding of interlayer current interdependence in a chiral device such as TBG and provides innovative tools for studying transport in complex systems.

In the first part of our study, we determined the four-probe resistance of TBG devices with four- and eight-layer differentiated contacts. In both configurations, our results reveal the emergence of longitudinal and Hall counterflow currents in the drag layer when current is injected into the drive layer. We observe that the magnitude of \( R_{\text{xx(drag)}} \) is comparable to that of \( R_{\text{xx(drive)}} \) only at low twist angles. Moreover, unlike the behavior observed for counterflow conductivity~\cite{Bistritzer11}, \( R_{\text{xx(drag)}} \) exhibits dips at energies corresponding to high DOS regions. In contrast, the Hall resistance $R_{\text{H(drag)}}$ remains of the same order of magnitude across different twist angles, in agreement with the observation of significant CD even at large twist angles \cite{Kim16,Zhai23}.

Our findings are robust against variations in current injection direction (armchair or zigzag), device-contact coupling, onsite disorder, lattice relaxation, and device size. While we do not include electron-electron interactions in our model, longitudinal and Hall counterflow effects are still expected in this regime. This expectation is supported by prior studies showing that, in bulk systems, electron-electron interactions primarily act to renormalize both longitudinal and transverse conductivities \cite{PhysRevB.102.125403,Calderon:2020aa,PhysRevB.109.155114}. Nevertheless, our results also reveal that the precise lineshape and magnitude of the resistances are highly sensitive to the nature of perturbations. With experimental electrical measurements of counterflow in mind, our analysis suggests that weak coupling between the contacts and the device provides the most favorable conditions. In this regime, the dipole Hall  can emerge clearly, characterized by the relation $R_{\text{H(drive)}} \approx -R_{\text{H(drag)}}$.

In the second part of our manuscript, we demonstrate that ML models can be integrated into the quantum transport algorithm to accelerate calculations. While this approach itself is not novel, the significant variations in conductance as a function of the twist angle allowed for the successful implementation of a novel divide-and-conquer strategy. Our predictive system, consisting of specialized GBR submodels for each cluster, enables more accurate and efficient predictions. However, the proposed ML model faced challenges in accurately predicting conductance for certain twist angles, highlighting the need for additional training data or more refined ML models to enhance precision. In this context, we also trained, validated, and tested Multi-Layer Perceptron (MLP) \cite{mlp1, mlp2, mlp3} and Support Vector Regression (SVR) \cite{svr1, svr2, svr3} for the nine clusters, but GBR demonstrated the best performance. The trained GBR model and the instructions to run it are available \cite{linkGBR}.

\section*{Acknowledgments}
M.H.G.K.  gratefully   acknowledges CAPES-PROSUC (88882.462011/2019-01) and  the high-performance computing cluster of the Mackenzie Presbyterian University (https://mackcloud.mackenzie.br).
D.A.B. acknowledges support from the Brazilian Nanocarbon Institute of Science and Technology (INCT/Nanocarbon), CAPES-PRINT (grants nos. 88887.310281/2018-00 and 88887.899997/2023-00), CNPq (309835/2021-6), and Mackpesquisa. D.A.B. also thanks Tobias Stauber and Guillermo G\'omez-Santos for bringing this topic to his attention and for their valuable feedback on the initial version of the manuscript.

\section*{DATA AVAILABILITY}
The data that support the findings of this article are openly available \cite{linkGBR}.

\bibliography{resistencia}

\end{document}